# Fermi Level Tuning of Epitaxial $Sb_2Te_3$ Thin Films on Graphene by Regulating Intrinsic Defects and Substrate Transfer Doping

Yeping Jiang,[1,2] Y. Y. Sun,[3] Mu Chen,[1,2] Yilin Wang,[1] Zhi Li,[1] Canli Song,[1,2] Ke He,[1] Lili Wang,[1] Xi Chen,[2] Qi-Kun Xue[1,2], Xucun Ma,[1*] and S. B. Zhang[3*]

[1]*Institute of Physics, Chinese Academy of Sciences, Beijing 100190, People's Republic of China*

[2]*State Key Lab of Low-Dimensional Physics, Department of Physics, Tsinghua University, Beijing 100084, People's Republic of China*

[3]*Department of Physics, Applied Physics, and Astronomy, Rensselaer Polytechnic Institute, Troy, New York 12180*

High-quality $Sb_2Te_3$ films are obtained by molecular beam epitaxy on graphene substrate and investigated by *in situ* scanning tunneling microscopy/spectroscopy. Intrinsic defects responsible for the natural p-type conductivity of $Sb_2Te_3$ are identified to be the Sb vacancies and $Sb_{Te}$ antisites in agreement with first-principles calculations. By minimizing defect densities, coupled with a transfer doping by the graphene substrate, the Fermi level of $Sb_2Te_3$ thin films can be tuned over the entire range of the bulk band gap. This establishes the necessary condition to explore topological insulator behaviors near the Dirac point.

[*] Corresponding authors. Email: xcma@aphy.iphy.ac.cn, zhangs9@rpi.edu

The discovery of three-dimensional (3D) topological insulators (TIs) in group V-chalcogenides, whose topological surface states (SS) host 2D helical Dirac fermions (DFs), has spurred tremendous interests in this class of materials as a potential vehicle for the exotic physical phenomena to exist only in DFs [1-6]. While $Bi_2Te_3$ and $Bi_2Se_3$ have been intensively studied in recent years [7-14], the study of $Sb_2Te_3$ in the context of its TI behavior is still rare [15-16]. While all these chalcogenide TIs have a simple SS band structure consisting of a single Dirac cone in the surface Brillouin zone, $Sb_2Te_3$ stands out for its unique advantages over $Bi_2Te_3$ and $Bi_2Se_3$, for example, the Dirac point lies well detached from the bulk band edges [16-17] to allow better measurement and manipulation of the DFs.

All available TIs are heavily populated by intrinsic defects. By introducing significant amount of free carriers, these defects pin the Fermi level to the bulk band edges to shadow electronic and spintronic responses of the DFs [5, 18-19]. As such, defect control for fine tuning the Fermi level inside bulk band gap, especially at the exact charge neutrality point, has become an important and challenging issue in TI studies [7-9, 20-23]. Previously, we established [24] a technique to control intrinsic defects in molecular beam epitaxy (MBE)-grown $Bi_2Te_3$ thin films by regulating substrate temperature. Both n-type and p-type $Bi_2Te_3$ can be obtained without extrinsic doping. Such a simple technique, however, does not work for $Bi_2Se_3$ (mostly n-type) and $Sb_2Te_3$ (mostly p-type). So far, p-type $Bi_2Se_3$ thin films can only be obtained by introducing high concentration of extrinsic dopants [8]. There has been no n-type $Sb_2Te_3$ reported unless one $\delta$-dopes $Sb_2Te_3$ by depositing a sub monolayer cesium on the surface of a thin film [16].

In this paper, we show that the difficulty with $Sb_2Te_3$ film is resolved by MBE growth of $Sb_2Te_3$ thin films on graphene substrate without the extrinsic δ layer. There exists a special growth temperature, $T_c$, at which the concentration of the holes reaches a minimum. By experimental determination of such a condition in conjunction with a transfer doping by the substrate, the Fermi level of $Sb_2Te_3$ can now be tuned to coincide with the Dirac point for films of five quintuple layers or thicker. Our combined scanning tunneling microscopy/spectroscopy (STM/STS) and density functional theory (DFT) calculation identify Sb vacancies ($V_{Sb}$) as the primary source of p-type conductivity for samples grown at low temperatures and Sb-on-Te antisites ($Sb_{Te}$) as that for samples grown at high temperatures. These findings are consistent with the observation of $T_c$ at which the total concentration of the two acceptor defects reaches a minimum.

Our experiment was performed under ultrahigh vacuum (base pressure < $1 \times 10^{-10}$ Torr) in a combined low-temperature STM and MBE chamber. An n-doped 6H-SiC(0001) substrate was used. The surface of the SiC, upon graphitization, was covered by a graphene bilayer [25]. During the growth, high-purity Sb and Te (both 99.9999%) were co-deposited onto the substrate from Knudsen cells. After the growth, the $Sb_2Te_3$ samples were transferred to an STM stage kept at 4.8 K, where the STM images and STS spectra ($dI/dV$) were taken. The STS spectra were acquired using a standard lock-in technique with a bias modulation of 1 $mV_{rms}$ at 987.5 Hz.

The mechanism for the MBE growth of $Sb_2Te_3$ is more complex than that of $Bi_2Se_3$ [25-26] and $Bi_2Te_3$ [24, 27]. During the growth of $Sb_2Te_3$, both Sb and Te fluxes are dominated by molecular species, mainly Sb tetramers ($Sb_4$) and Te dimers ($Te_2$). This is different from growing $Bi_2Se_3$ and $Bi_2Te_3$, where Bi is mainly in the atomic form. A high substrate temperature and a low beam flux are required to avoid Sb clustering and the formation of structural defects due to the low mobility of $Sb_4$. In addition, the sticking coefficient of the $Sb_4$ on graphene is lower than that on $Sb_2Te_3$. At a fixed beam flux, the substrate temperature must be lower than the threshold $T_0$ for $Sb_2Te_3$ nucleation. After the nucleation stage, however, the temperature can be increased to improve the film quality. Thus, a two-step procedure was used. The Knudsen cell temperatures, $T_{Sb}$ and $T_{Te}$, were kept at 330 and 225 °C, respectively, which yield a Te-rich growth condition with a nominal Te/Sb flux ratio ($\theta$) of ~10 and a low growth rate of ~0.2 quintuple layers (QL) per minute. At this beam flux, $T_0$ is ~190 °C. The two-step growth temperatures $T_1$ and $T_2$ are thus set at ~190 and 200-250 °C, respectively.

Figure 1(a) shows a typical STM image of the $Sb_2Te_3$ film. A regular terrace step of ~1.01 nm or 1 QL clearly indicates a layer-by-layer growth mode. The terraces are atomically flat and of good crystallinity, as evidenced by the atomic resolution image (inset of Fig. 1(a)). The (111) lattice constant is measured to be ~0.426 nm. The layer-by-layer growth mode is further confirmed by the evolution of the quantum well states (QWS) in the thin film. This can be seen in Fig. 1(b) where a series of *dI/dV* spectra are taken on films of 1 - 8 QL. Within the Tersoff-Hamann approximation [28], *dI/dV* spectra correspond to the local density of states (LDOS) for the film. The peaks on the low-energy side of the LDOS (marked by arrows) correspond to the QWS in the bulk-like valence band (BVB), whereas the steps in the high-energy side correspond to the QWS in the bulk-like conduction band (BCB). The separation between the QWS decreases with increasing film thickness. The energy difference between the highest BVB QWS and the lowest BCB QWS is related to the bulk energy gap. It decreases from ~900 meV at 1 QL to ~400 meV at 8 QL, as can be seen in Fig. 1(b). More details about characterizing the QWS states by STS is provided in the Supplementary Information.

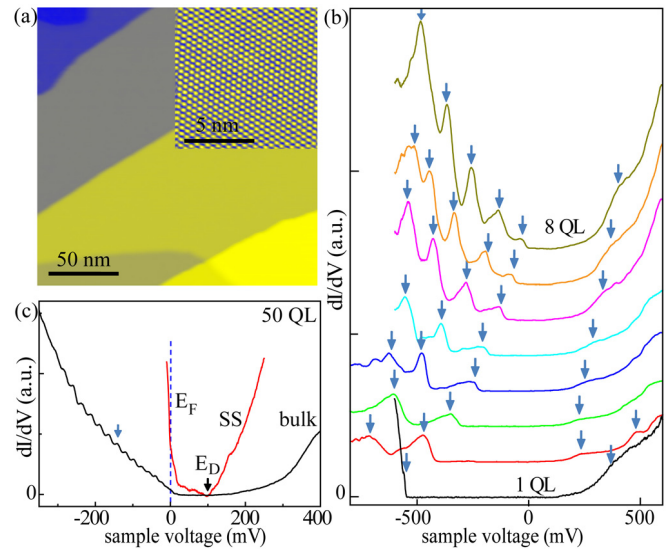

FIG. 1. (color online) (a) STM image of a $Sb_2Te_3$ film less than 10 QL ($V_{bias}$ = 5.0 V, $I$ = 50 pA). Inset is an atomic resolution image (-1.0 V, 50 pA). Bright spots correspond to surface Te. (b) *dI/dV* spectra (0.3 V, 50 pA) of films from 1 to 8 QL. Arrows indicate the energies of bulk-like QWS. (c) *dI/dV* spectra of bulk (0.3 V, 50 pA) and SS (0.25 V, 200 pA) on a 50 QL $Sb_2Te_3$ film.

For a thick film (~50 QL), the LDOS spectrum in Fig. 1(c) shows a bulk gap of

~300 meV. The oscillations due to QWS in the BVB can still be clearly observed with a peak-to-peak distance of ~23 meV. In the energy range of the bulk gap, the SS can be probed, showing a V-shaped spectrum with the zero-conductance point at ~100 meV. The zero-conductance point corresponds to the Dirac point ($E_D$) of the SS [17]. The position of $E_D$ (~100 meV in Fig. 1(c)) is insensitive to the growth parameters for thick films, for which the Fermi level is pinned around the BVB edge. This result reinforces the notion that both $Sb_2Te_3$ films and bulk samples are p-type [15-16], which has hindered the use of photoemission techniques, a major tool for the investigation of the physical properties of the TIs [7, 9].

Five types of defects under various growth conditions are observed. Figures 2(a) and (b) show the STM images for samples grown at ~190 and ~230 °C, respectively, with $\theta$ = ~10. Four types of defects (labeled I, II, IV, and V) can be seen. At even more Te-rich condition (e.g., $\theta$ = ~20), another type of defects (labeled III) is observed, as shown in Fig. 2(c). Defects I, II, IV, and V are all acceptors, as judged by the Dirac point position in the corresponding LDOS spectra (see Supplementary Materials Fig. S3). Figure S2 in the Supplementary Information explains how to measure the Dirac point using the STS.

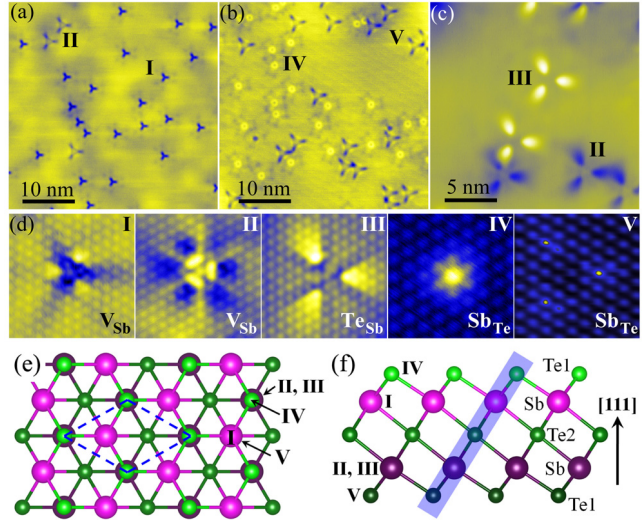

FIG. 2. (color online) (a)-(c) Large-area STM images for various defects, labeled from I to V. Tunneling conditions: (a) 1.0 V, 50 pA; (b) 1.0 V, 50 pA; and (c) 0.4 V, 50 pA. (d) High resolution STM images for defects I to V. Tunneling conditions: (I) 0.2 V, 100 pA; (II) 0.1 V, 100 pA; (III) 0.2 V, 100 pA; (IV) -0.5 V, 50 pA; and (V) -1.0 V, 50 pA. (e) and (f) Top and side views of $Sb_2Te_3(111)$ showing positions of the defects. Dashed frame is the surface unit cell. Shaded region is a Te1-Sb-Te2-Sb-Te1 chain in a single QL. Te and Sb atoms are denoted by green (small) and pink (large) spheres, respectively. Note that, looking from top, I is above V and IV is above II and III (see Fig. S1 for more information).

Figure 2(d) shows high resolution STM images for the observed defects. The corresponding positions of these defects, as shown in Figs. 2(e) and 2(f), are assigned by measuring their lateral registries with respect to the top layer Te atoms and by considering the spatial distribution of their STM features. The chemical bonding in $Sb_2Te_3$ is similar to that in $Bi_2Te_3$ and $Bi_2Se_3$ [10] and can be approximated by strongly interacting $pp\sigma$ chains of atomic $p$ orbitals [29] with the order

Te1-Sb-Te2-Sb-Te1, as shown in Fig. 2(f). A defect is expected to perturb the electronic states predominantly along three equivalent, 120°-apart *ppσ* chains passing the defect, resulting in three spots at surface atoms terminating the chains. The center joining the three spots is the location of the defect. Surface Te-site defect (i.e., defect IV) is an exception for which only one spot should be seen. Using this argument, we identify that defect I is centered on a Sb site in the second layer, defects II and III are on Sb sites in the fourth layer, and defects IV and V are on Te sites in the first and fifth layers, respectively. In Fig. 2(d), defects I, II, IV, and V show depression at positive bias and protrusion at negative bias, implying that they are electron acceptors, in accordance with the LDOS measurement using STS (Fig. S3(c)). On the other hand, defect III shows opposite contrast, for example, to defect II (see Fig. 2(c)) even though they are on the same atomic site, implying that defect III is an electron donor. Thus, we may attribute defects I and II to Sb vacancies ($V_{Sb}$), defect III to Te-on-Sb antisite ($Te_{Sb}$), and defects IV and V to $Sb_{Te1}$ in different atomic layers, as illustrated in Fig. 2(f).

The occurrence of these defects is intimately related to growth temperature *T* and flux *θ*: $V_{Sb}$ only exists at relatively low *T* and disappears above ~200 °C. $Te_{Sb}$ coexists with $V_{Sb}$ at low *T* (< 200 °C) and high *θ*. $Sb_{Te1}$ appears at ~200 °C. $Sb_{Te2}$ has not been observed for 190 °C < *T* < 250 °C, indicating that it has higher formation energy than $Sb_{Te1}$.

The defect assignments above are verified by DFT calculations, performed using the VASP code [30]. The generalized gradient approximation [31] was used for the exchange-correlation functional. Core electrons were represented by the projector augmented wave potentials [32]. Plane waves with a cutoff energy of 250 eV were used as the basis set. A (4×4×1) 240-atom supercell with lattice constants ($a$ = 4.264 Å and $c$ = 30.458 Å) was used to model the defects. A single special $k$-point at (7/24, 1/12, 1/4) was used to sample the Brillouin zone. Defect formation energies were calculated following the formalism in Ref. [33].

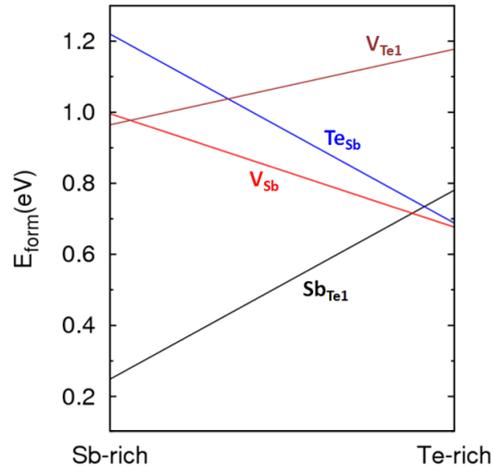

FIG. 3. (color online) Calculated intrinsic defect formation energy under various growth conditions: the Te-rich condition corresponds to the chemical potential of Te ($\mu_{Te}$) set to that of bulk Te, whereas the Sb-rich condition corresponds to $\mu_{Sb}$ set to that of bulk Sb with the constraint $2\mu_{Sb} + 3\mu_{Te}$ equal to the total energy per formula bulk $Sb_2Te_3$.

Figure 3 shows the calculated formation energy of possible intrinsic defects in $Sb_2Te_3$. At a strong Te-rich growth condition, $V_{Sb}$ is the lowest-energy defect. When the system becomes less Te-rich, which could be a result of increased substrate temperature [24], $Sb_{Te1}$ becomes the lowest-energy defect in agreement with experiment. Our calculations confirmed that both defects are acceptors. The lack of $Sb_{Te2}$ in our experiment can be explained because it is less stable than $Sb_{Te1}$ by 0.2 eV. The lack of $V_{Te}$ can also be explained by its high formation energy. In addition, at a highly Te-rich condition corresponding to higher $\theta$, $Te_{Sb}$ is only slightly higher in energy than $V_{Sb}$, in line with the coexistence of $Te_{Sb}$ with $V_{Sb}$ in Fig. 2(c).

We are unable to obtain $Sb_2Te_3$ samples in which the $Te_{Sb}$ donor dominates. A too high $\theta$ would degrade the sample morphology. To compensate for intrinsic acceptors, carriers of opposite sign must be introduced. In this regard, the n-doped graphene substrate can serve as an electron donor. Figure 4(a) shows the thickness dependence of $E_D$ with respect to $E_F$. Between 4 QL and 5 QL, the two levels cross, implying an n-to-p type conversion. The substrate transfer doping effect decays with film thickness with the critical thickness at about ~8 QL for this set of samples, above which $E_D$ approaches the bulk value.

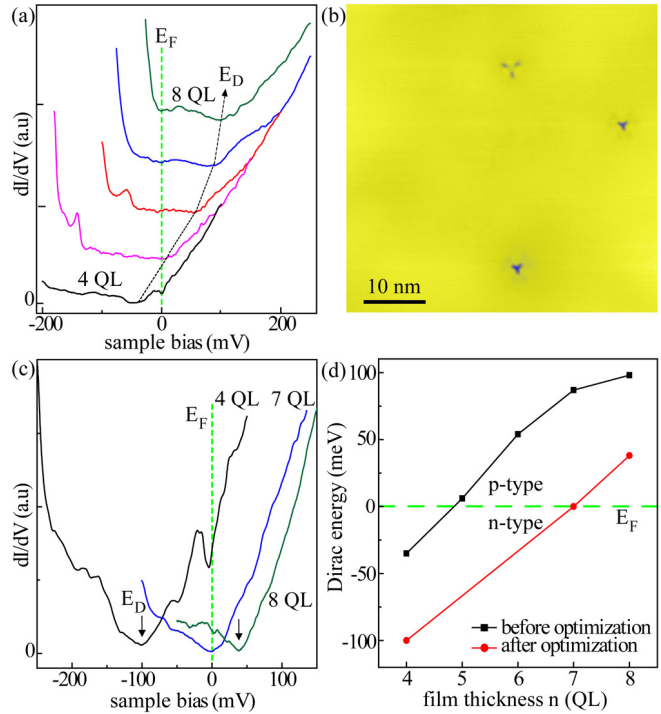

FIG. 4. (color online) (a) Thickness dependent SS spectra of $Sb_2Te_3$ films from 4 to 8

QL. The spectra are shifted vertically for clarity. (b) STM image (1.0 V, 50 pA) of the optimized sample. (c) SS spectra for 4, 7, and 8 QL optimized films. (d) Evolution of $E_D$ with film thickness.

With the knowledge on the defect formation, the thickness of the n-type region can be further increased by reducing the density of intrinsic acceptors to enable a wider range of Fermi level tuning. Figure 3 predicts that at the transition region between $V_{Sb}$ and $Sb_{Te1}$, the acceptor density is the lowest. Experimentally, we determined that this corresponds to the growth condition: $\theta =$ ~10 and $T_2 =$ ~200 °C, at which the amount of intrinsic defects is greatly reduced. Figure 4(b) shows that the defect density is on the order of ~$1.2 \times 10^{11}$/cm$^2$, compared to, for example, ~$1.7 \times 10^{12}$/cm$^2$ for other samples. Accordingly, Fig. 4(d) shows consistently that $E_D$ is shifted downwards by ~60 meV. In particular, $E_D$ is around -100 meV at 4 QL; $E_D$ nearly coincides with $E_F$ at 7 QL; $E_D$ is above $E_F$ at 8 QL as shown in Fig. 4(c), signaling the transition to p-type region. These results demonstrate that $E_F$ can be controlled in the energy range of ±100 meV around $E_D$ in our MBE samples while minimizing the effects of intrinsic or extrinsic defects.

In summary, by a combined STM/STS experiment and DFT calculation, we identify major intrinsic defects in MBE-grown $Sb_2Te_3$ on graphene substrate and explain the p-type behavior of the undoped films. We show that, by combining substrate n-doping with intrinsic defect control, one could tune the Fermi level in $Sb_2Te_3$ thin films to across nearly the whole band gap region of bulk $Sb_2Te_3$. This paves the way for further study of the TI behavior in large-gap $Sb_2Te_3$: for example, due to the helical nature of TI SS, the spin direction with respect to the momentum changes sign for states below and above $E_D$ [34]. Thus, the Fermi surface of the SS may undergo a spin-texture reversion at where the n-p transition takes place. In films with both electron and hole pockets of the DFs, exciton condensation may also be observed [4].

Work in China was supported by the National Science Foundation and Ministry of Science and Technology of China. All STM topographic images were processed by the WSxM software. Work at RPI was supported by the U.S. Department of Energy under Grant No. DE-SC0002623. Supercomputer time was provided by the National Energy Research Scientific Computing Center, supported by the Office of Science of the U.S. Department of Energy under Contract No. DE-AC02-05CH11231.

# Fermi Level Tuning of Epitaxial Sb$_2$Te$_3$ Thin Films on Graphene by Regulating Intrinsic Defects and Substrate Transfer Doping

（Supplemental Information）

Yeping Jiang,[1,2] Y. Y. Sun,[3] Mu Chen,[1,2] Yilin Wang,[1] Zhi Li,[1] Canli Song,[1,2] Ke He,[1] Lili Wang,[1] Xi Chen,[2] Qi-Kun Xue,[1,2] Xucun Ma,[1*] and S. B. Zhang[3*]

[1]Institute of Physics, Chinese Academy of Sciences, Beijing 100190, People's Republic of China

[2]State Key Lab of Low-Dimensional Physics, Department of Physics, Tsinghua University, Beijing 100084, People's Republic of China

[3]Department of Physics, Applied Physics, and Astronomy, Rensselaer Polytechnic Institute, Troy, New York 12180

## I. Identification of defect positions based on high resolution STM images and crystal structure of Sb$_2$Te$_3$

Sb$_2$Te$_3$ has the rhombohedral crystal structure [1] as shown in Fig. S1a, with a layered structure stacked in the [111] direction (or z axis) and a triangle lattice in the xy plane. In the xy plane, the lattice constant is 0.426 nm as measured by atomically resolved STM images of the (111) Te surface (Fig. 1a). The sequential five atomic layers Te1-Sb-Te2-Sb-Te1 are called as a quintuple layer (QL). 1 QL is measured to be about 1.01 nm by STM. Like Bi$_2$Se$_3$ and Bi$_2$Te$_3$, within one QL, the p orbitals of Sb and Te atoms form σ-bonded chains between the atomic layers [2], while between the QLs, the atom-atom interactions are Van der Waals force in nature. The Te1(111) is the natural cleavage surface. The Te atoms in the third Te layer below the surface are denoted as Te3, and the first and second Sb atomic layer are indicated as Sb1 and Sb2, respectively (Fig. S1b). The stack sequence of QLs is ABCABC…, so the period in the z direction is 3 QL. In the MBE growth, Sb$_2$Te$_3$ films follow the layer-by-layer growth mode.

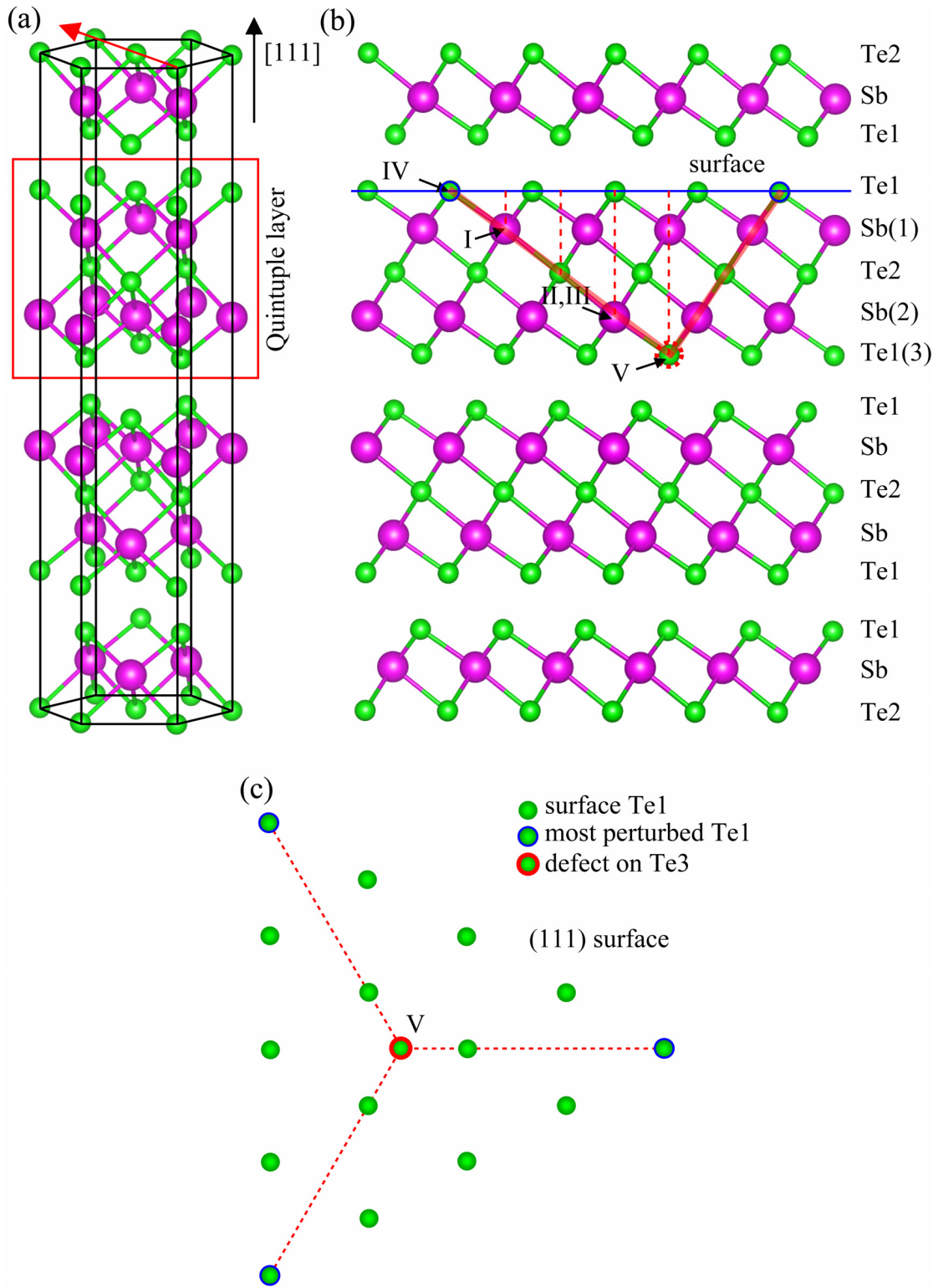

FIG. S1. Crystal structure. (a) Crystal structure of Sb$_2$Te$_3$. One quintuple layer with atomic layer sequence Te1-Sb-Te2-Sb-Te1 is indicated by the red square. (b) Side view of the structure in the direction indicated by a red arrow in (a) and parallel to the quintuple layer. (c) Schematic of most perturbed surface atoms in the presence of a defect on Te3.

The ppσ bonding scheme is briefly shown in the side-view crystal structure in Fig. S1b. In the presence of a defect, vacant or substitutive, the electronic states along the chain will be perturbed most. The red lines in Fig. S1b connect the defect (on Te3) and the most-perturbed surface Te atoms. The structure of $Sb_2Te_3$ has threefold symmetry, so there will be three Te atoms which may be most-perturbed in their electronic structure. The STM images are intimately related to the electronic density of states (DOS), hence the defects will have strong perturbation on the images. The structure of perturbation will have a triangle form, and the distance between the most-perturbed three atoms is indicative of which layer the defect lies in. For example, as shown in the diagram of Fig. S1c, substitutional defect $Sb_{Te}$ result in three spots separated by 4 atoms in the high-resolution STM image (defect V in Fig. 2d). Furthermore, the lateral position of the center of the perturbation provides another evidence for the defect identification. For example, Sb2 and surface Te atoms share the same lateral positions, as seen in the [111] direction (Fig. 2e and Fig. S1b). So the defects on Sb2 will have perturbed electronic structure centering at surface Te positions in the highly-resolved STM image (defect II, III as shown in Fig. 2d). The defects in the deeper QL can be hardly seen in the STM images. The perturbation on surface Te1 by defects on Te3 is already very weak.

**II. dI/dV spectra taken on $Sb_2Te_3$ thin films and their correspondence to the band structure**

In the thin film case, the gapped bulk states of topological insulator $Sb_2Te_3$ are quasi 3D because of the confinement in z direction. $Sb_2Te_3$ has an indirect bulk gap, with valence band maximum not being at Γ [1]. The lower panel of Fig. S2 is a schematic diagram of the band structure of the $Sb_2Te_3$ (111) surface. The bulk valence band (BVB) and bulk conduction band (BCB) are quantized, in the gap of which the surface states (SS) reside. The SS are 2D massless Dirac fermions and form a Dirac cone as indicated in the diagram. The crossing of two surface bands connected by time reversal symmetry is the Dirac point.

The STM dI/dV spectrum measures the surface density of states (DOS), in which the topological SS are much smaller than the projected bulk (or bulk-like) states. As shown in Fig. 1b, a gap feature appears in the dI/dV spectra of $Sb_2Te_3$ thin films, and it becomes larger in thinner films (about 800 meV and 500 meV in the 1 QL and 4 QL cases, respectively), extending the SS energy range and providing a good way of band engineering. Figure S2 shows the dI/dV spectrum of a 13 QL $Sb_2Te_3$ film with a bulk gap of about 300 meV. In the BVB side, the quantum well states (QWS) can be clearly seen. The M-shaped BVB bands may have Van Hove singularities in the DOS at energies corresponding to the top of 'M'. This may explain the observed well-defined BVB QWS in the dI/dV spectra. The dashed lines in Fig. S2 illustrate this correspondence. The issue about the effect of M-shaped band on the DOS has been addressed in a previous work on Pb films by using STM combined with ARPES [3]. The lack of sharp BCB edge or QWS may be due to the fact that BCB states are nearly parabolic, while BVB states are M-shaped. The quantized parabolic BCB states will result in stepped DOS feature which may only be seen when these quantized states are well separated, as shown in the case of very thin films (Fig. 1b).

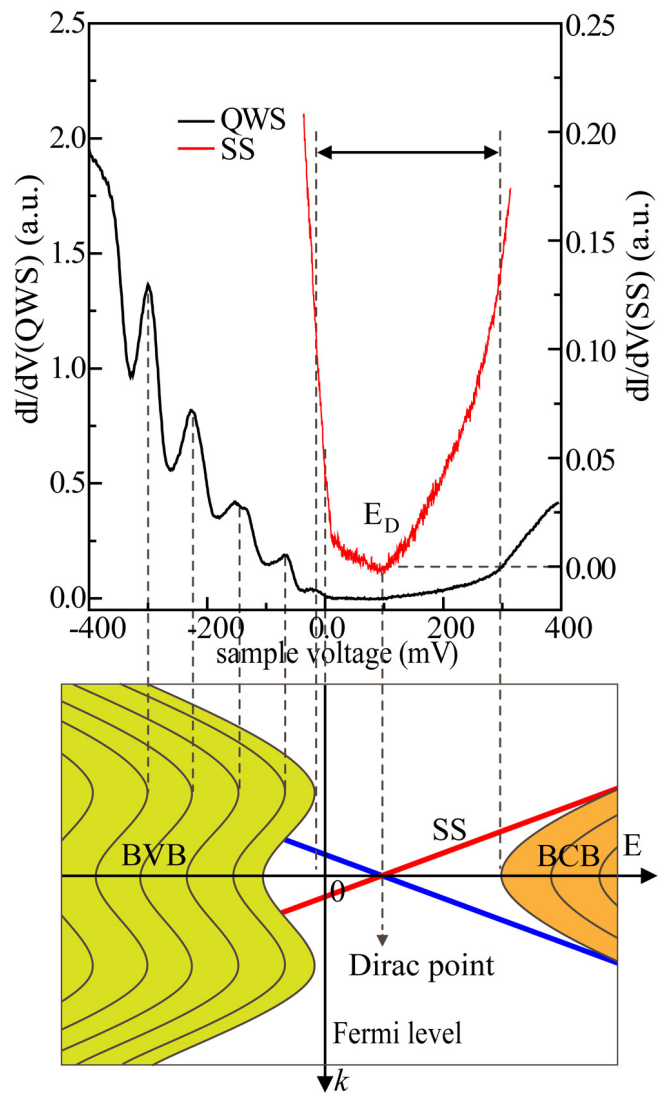

FIG. S2. STM dI/dV spectra of QWS (black) and SS (red) on 13 QL $Sb_2Te_3$ thin films. Gap conditions: 0.4 V, 100 pA for QWS; 0.3 V, 200 pA for SS. The lower panel is a schematic of surface band structure of $Sb_2Te_3$ thin films near Γ.

Because of much lower DOS of SS, by setting a closer tunneling gap, the detailed SS feature can be probe in the energy range of bulk gap. In the SS spectrum (red curve), the nearly zero conductance can be attributed to the Dirac point ($E_D$), about 100 meV above the Fermi level. This zero conductance is a proof that the Dirac point of $Sb_2Te_3$ is well separated from the bulk states. About 90 meV below $E_D$, a sharp increase in DOS is an indication of bulk states (85 meV for 50 QL films, as shown in Fig. 1c), or the edge of the first BVB QWS. In most samples without defect control, the Fermi level is pinned near the BVB edge, as shown in Fig. S2 and Fig. 1c.

**III. dI/dV spectra taken on $Sb_2Te_3$ thin films populated by different defect types**

The SS spectra shown in Fig. S3a-c were taken on films populated by $V_{Sb}$, $Sb_{Te}$, and $V_{Sb}+Te_{Sb}$, respectively. The Dirac points are about 100 meV, 105 meV, and 80 meV above the Fermi level. The defects $V_{Sb}$, $Sb_{Te}$ are thus proved to be p-type. $Te_{Sb}$ has opposite contrast compared with $V_{Sb}$ in the STM images, implying its n-type nature. This point is supported by the fact that samples populated by both $V_{Sb}$ and $Te_{Sb}$ are less p-doped than those only having $V_{Sb}$ (Fig. S3a and S3c).

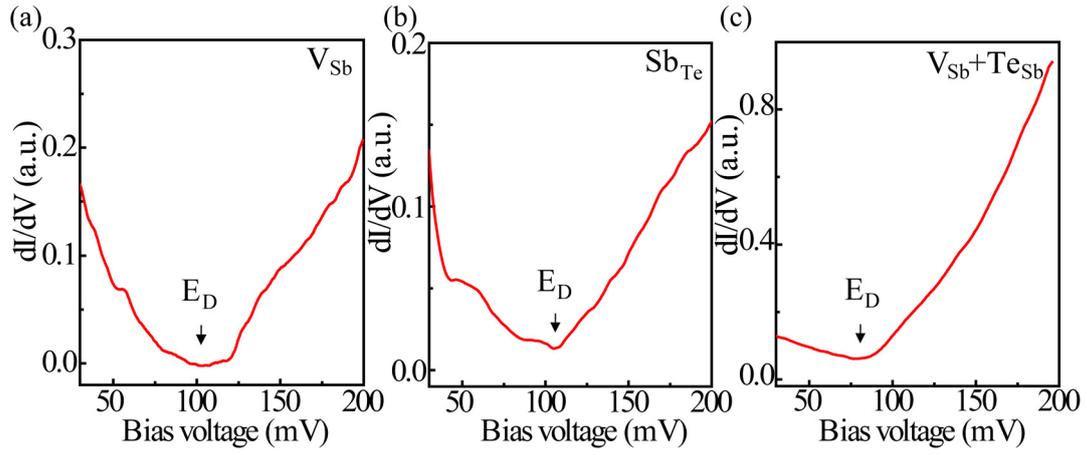

FIG. S3. Corresponding SS spectra (0.25 V, 200 pA) for samples in Fig. 2a-c. The Dirac points are indicated by arrows as determined in the way shown in Fig. S2.